\documentstyle[aps,prl]{revtex}
\begin{document}

\twocolumn[\hsize\textwidth\columnwidth\hsize\csname   
@twocolumnfalse\endcsname   

\title{Wetting layer thickness and early evolution of epitaxially
strained thin films}
\author{Helen R. Eisenberg\cite{email1} and Daniel Kandel\cite{email2}}
\address{Department of Physics of Complex Systems,\\
Weizmann Institute of Science, Rehovot 76100, Israel}

\maketitle

\begin{abstract}
We propose a physical model which explains the existence of finite
thickness wetting layers in epitaxially strained films. The finite
wetting layer is shown to be stable due to the variation of the
nonlinear elastic free energy with film thickness. We show
that anisotropic surface tension gives rise to a metastable enlarged
wetting layer. The perturbation amplitude needed to
destabilize this wetting layer decreases with increasing lattice
mismatch. We observe the development of faceted islands in unstable
films.
\end{abstract}
\vspace{0.3cm}
{\hspace*{2.cm}}PACS numbers: 68.55.Jk, 68.35.-p, 81.15.Aa
\vspace{0.3cm}
] 
\input epsf 

Epitaxial deposition of a thin film onto a substrate in cases involving
lattice mismatch is central in the fabrication of semiconductor and
optoelectronic devices. The lattice mismatch between the substrate and
the
film generates strain in the deposited film, which can cause film
instability
unfavorable to uniform flat film growth. The strained film can relax
either
by the introduction of dislocations or by the formation of
dislocation-free
islands on the film surface via surface diffusion. Early film growth
tends
to occur via the second mechanism and we shall only consider dislocation
free films. It has been observed experimentally \cite
{mo,floro} that dislocation free flat films of
less
than a certain thickness (the critical wetting layer) are stable to
surface
perturbations, while thicker films are unstable. The thickness of the
wetting layer is substance dependent and decreases with increasing
lattice
mismatch strain \cite{floro}, $\varepsilon=(a_{s}-a_{f})/a_{f}$, where
$a_{s}$
and $a_f$ are the substrate and film lattice constants. Above the
critical wetting layer 3D dislocation-free islands form. Prediction and
control of wetting layer thickness and an understanding of early thin
film evolution are important for the improved fabrication of
semiconductor devices.

Despite considerable efforts (see, e.g.,
\cite{tersoff,spencer2,chiu,kukta,daruka,spencer}),
the physics of the critical wetting layer is poorly understood, and the
purpose of this Letter is to study its properties by considering the
following two important questions: First, why is there a critical,
stable
wetting layer and what controls its thickness? Second, since in most
cases
heteroepitaxial growth is done below the roughening
transition, how does
anisotropic surface tension affect the thickness of the critical wetting
layer? Here we attempt to answer these questions and to study
early
film evolution without deposition. A later paper will look at the
effects of
deposition and long-term growth.

We studied an elastically isotropic system under plane strain
\cite{timoshenko}, 
which causes the system to be effectively two dimensional. 
The surface of the solid is at $y=h(x,t)$ and the film is in the $y>0$
region with the film-substrate
interface at $y=0$. The system is invariant in the $z$ direction and all
quantities are calculated for a section of unit width in the $z$
direction.

We assume that surface diffusion is the dominant mass transport
mechanism, leading to the following evolution of the surface profile
\cite{mullins57}: 
\begin{equation}
\frac{\partial h(x,t)}{\partial t}=K\frac{\partial ^{2}}{\partial x
^{2}}\frac{\delta F}{\delta h(x,t)}~,
\label{evol}
\end{equation}
where $F$ is the free energy of the system and $K$ is a positive
constant. The
free energy is composed of elastic and surface terms: 
\begin{equation}
F=F_{el}+\int dx~\gamma\sqrt{1+(\partial h/\partial x)^2}~,
\label{free energy}
\end{equation}
where $\gamma$ is the surface
tension and $F_{el}$ is the elastic free energy, which includes also any
elastic contributions to the surface tension. We express $F_{el}$ as
$F_{el}=F_{el}^{(0)}+\delta F_{el}$, where $F_{el}^{(0)}$ is the elastic
free energy of the zero strain reference state, and $\delta F_{el}$ is
calculated from linear elasticity theory. For each value
of $x$, the reference state corresponds locally to a {\em flat} film of
thickness $h(x)$; i.e., $F_{el}^{(0)}=\int dx\int_{-\infty}^{h(x)} dy
f_v^{(0)}(h(x),y)$, where
$f_v^{(0)}(h(x),y)$ is the elastic free energy per unit volume of a flat
film of thickness $h(x)$. 

Due to the lateral variations in this
reference state, the reference stress does not satisfy the condition of
mechanical equilibrium. However, the necessary corrections vanish in the
limit $a/\lambda\rightarrow 0$, where $a$ is the length scale over which
stress varies in the $y$ direction and $\lambda$ is the lateral length
of typical surface structures. This is because in this limit there are
no lateral variations in the
reference stress. As typical experimental islands have $\lambda\sim
100\;nm$, and as $a$ is of the order of the lattice constant (see
below), the corrections to the reference stress are small and have been
ignored.   

For convenience we work in terms of the
reference elastic free energy per unit length in the $x$ direction, 
$f_{el}^{(0)}(h(x))\equiv\int_{-\infty}^{h(x)} dy f_v(h(x),y)$, instead
of the free energy per unit volume. As discussed below, the dependence
of $f_{el}^{(0)}$ on $h$ is a nonlinear phenomenon and cannot be
calculated from linear elasticity theory.

For small strains, the stress is linear in the strain, i.e., 
$\sigma _{ij}=s_{ij}^{m}+c_{ijkl}e_{kl}$, where
repeated indices
are summed over. Here $e_{ij}$ is the strain tensor, $\sigma _{ij}$ is
the total stress tensor, $s_{ij}^{m}$ is the stress in the zero strain
reference state due to the lattice mismatch and $c_{ijkl}$ are the
elastic coefficients of the material.
According to linear elasticity theory, $\delta
F_{el}=\int dx\int_{-\infty }^{h(x)}dy\left(
s_{ij}^{m}e_{ij}+\frac{1}{2}c_{ijkl}e_{ij}e_{kl}\right)$. In terms of
the stress tensor, we
find 
\begin{eqnarray}
F_{el} &=&\int dx~f_{el}^{(0)}  \nonumber \\
&+&\int dx\int_{-\infty }^{h(x)}dy\left( \frac{1}{2}S_{ijkl}\sigma
_{ij}\sigma _{kl}-\frac{1}{2}S_{ijkl}s_{ij}^{m}s_{kl}^{m}\right) ~,
\label{fel}
\end{eqnarray}
where we have used the inverted Hooke's law $e_{ij}=S_{ijkl}s_{kl}$.
$S_{ijkl}$ are the compliance coefficients of the material. Combining
Eqs.\
(\ref{free energy}) and (\ref{fel}), we arrive at an
expression for $\delta F/\delta h$ at the surface: 
\begin{equation}
\frac{\delta F}{\delta h} =\left( \widetilde{\gamma }\kappa
+\frac{df_{el}^{(0)}}{dh}+\frac{1}{2}%
S_{ijkl}\sigma _{ij}\sigma
_{kl}-\frac{1}{2}S_{ijkl}s_{ij}^{m}s_{kl}^{m}\right) \Omega ~,
\label{linear chem pot}
\end{equation}
where $\Omega$ is the atomic area of the solid, $\kappa$ is the
surface curvature, $\widetilde{\gamma }(\theta)=\gamma
(\theta)+{\partial^{2}\gamma}/{\partial\theta^{2}}$ is the surface
stiffness and $\theta$
is the angle between the normal to the surface and the $y$ direction.
As Eq.\ (\ref{linear chem pot}) gives $\delta F/\delta h$ at the solid
surface, all variables in the equation are also given at the surface.
Both $df_{el}^{(0)}/dh$
and $s_{ij}^{m}$ must vanish when $h\leq 0$, since then the film is
absent.
In principle, Eq.\ (\ref{linear chem pot}) should also contain
derivatives of $\gamma$ with respect to $h$. However, we
believe that the variation of surface tension with $h$ away from a step
dependence is due to elastic effects. Since we included all elastic
contributions in the zero-strain elastic free energy, we modeled
$\gamma$ as a step function, taking the value of the substrate surface
tension for $h\leq0$ and the film surface tension for $h>0$. Thus all
partial derivatives
of $\gamma$ with respect to surface height vanish and were omitted from
Eq.\ (\ref{linear chem pot}).

Equations (\ref{evol}) and (\ref{linear chem pot}) form a complete model
of
film evolution. In order to solve this model one has to evaluate $\gamma
(\theta)$, 
$S_{ijkl}$, $f_{el}^{(0)}$ and $s_{ij}^{m}$. The first two are material
properties, while the last two are properties of the reference state,
from
which one can also calculate the stress tensor $\sigma _{ij}$ using
linear
elasticity theory. Before estimating these quantities we present the
results
of the linear stability analysis of an isotropic flat film of
thickness $C$. The analysis was carried out using a method similar to
those
used in \cite{asaro} for an infinite film. The height of the
perturbed film takes the form $%
h(x,t)=C+\delta (t)\sin kx$. We assumed that the force on the surface
due to
surface tension is negligible compared to the force due to mismatch
stress,
and that the stress $\sigma _{ij}$ vanishes deep in the substrate. Using
linear elasticity theory and Eq. (\ref{linear chem pot}) with
these assumptions, we calculated $\delta F/\delta h$ to first order in
the perturbation, and combined the results
with
the general evolution equation (\ref{evol}) to obtain the following
equation
for the evolution of $\delta (t)$: 
\begin{equation}
\frac{d\delta }{dt}=K\left[ -k^{4}\widetilde{\gamma }_{0}-k^{2}
\frac{
d^{2}f_{el}^{(0)}}{dh^{2}}
+2k^{3}\frac{\eta^2(h)}{M}\right]_{h=C} \delta ~,  \label{lin surface
evolution}
\end{equation}
where $\widetilde{\gamma }_{0}\equiv \widetilde{\gamma }(\theta =0)$,
$M$ is
the plain strain modulus derived from the elastic constants of the
isotropic
material, and $\eta(h)$ is $s_{xx}^m$ at the surface of a flat
film of thickness $h$. $s_{xy}^m$ vanishes because the flat film is
hydrostatically strained, and $s_{yy}^m=0$ since in the reference state
the force on the surface in the $y$ direction vanishes. 

Equation (\ref{lin surface evolution}) implies that the flat film is
stable at all perturbation wavelengths as long as 
\begin{equation}
\frac{\left[ \eta(C)\right] ^{4}}{M^{2}}\leq \widetilde{\gamma
}_{0}\left. 
\frac{d^{2}f_{el}^{(0)}}{dh^{2}}\right| _{h=C}~,  \label{crit}
\end{equation}
and the equality holds at the critical wetting layer thickness.
$\widetilde{\gamma }_{0}$ is positive if $\theta =0$ is a surface seen
in the equilibrium free crystal \cite{herring}. At a perfect facet,
$\widetilde{\gamma }_{0}\rightarrow \infty $. Hence, as $\theta =0$ is a
facet direction for most of the materials used in epitaxial films,
$\widetilde{\gamma}_{0}$
is large and positive. Therefore, a linearly stable wetting layer of
finite
thickness can exist only if $d^{2}f_{el}^{(0)}/dh^{2}>0$. Note that
$\eta$
depends linearly on the lattice mismatch $\varepsilon $, and hence the
l.h.s.\
of (\ref{crit}) is proportional to $\varepsilon ^{4}$, while the r.h.s.\
of (\ref{crit}) is proportional to $\varepsilon ^{2}$ due to the
dependence
of $f_{el}^{(0)}$ on lattice mismatch. Therefore, if
$d^{2}f_{el}^{(0)}/dh^{2}>0$, the thickness of the wetting layer
increases with decreasing lattice
mismatch and diverges in the limit $\varepsilon \rightarrow 0$.

Having recognized the importance of the elastic free energy of the
reference
state, $f_{el}^{(0)}$, and its dependence on film thickness, we now turn
to
estimate it. This free energy depends strongly on the mismatch stress
$s_{ij}^{m}$, and its dependence on the $y$ coordinate. As a result of
the
sharp interface between the substrate and the film, we expect
$s_{ij}^{m}$ to
behave as a step function of y with small corrections due to elastic
relaxation. If we ignore these small corrections, the resulting free
energy $f_{el}^{(0)}$, is proportional to film thickness, and its second
derivative vanishes. Hence, according to Eq.\ (\ref{crit}), the
thickness of the critical wetting layer vanishes. The correction due to
elastic relaxation is therefore extremely important. It turns out that
this correction vanishes within linear elasticity theory.
This led some investigators \cite{kukta} to claim that the
variation in free energy over the
interface was due to non-elastic effects, e.g.\ film/substrate material
mixing
over the interface. However, we claim that this is not necessary, since
nonlinear elasticity can explain the corrections to the step-function
form
of the free energy.

Ideally, first principles, substance-specific calculations should be
performed in order to evaluate $\eta(h)$ and $f_{el}^{(0)}(h)$, and we
intend to carry out such calculations in the future. However, the
qualitative general behavior of $f_{el}^{(0)}(h)$ can be obtained from
much
simpler models. To demonstrate this point we carried out the calculation
for two dimensional networks of balls and springs of varying lattice
type and spring constants. In this model the
balls are connected by
springs which obey Hooke's law. The natural spring length had a step
variation over the interface. The film underwent a hydrostatic
transformation strain so that its lattice constant became that of the
substrate. The network was then allowed to relax whilst being
constrained in the $x$ direction and free in the $y$ direction, so that
the system boundaries in the $x$ direction were fixed to the natural
substrate length. 

We calculated the mismatch stress within
the
film and at the film surface for films of varying
thickness. However, we decided to use the step function form of mismatch
stress, $\eta(h>0)=M\varepsilon$, where $M\varepsilon $ is the mismatch
stress in
an infinite film, as variations in $\eta$ only slightly altered the
wetting layer
thickness predicted from (\ref{crit}).  
We also calculated the nonlinear elastic free energy of the relaxed
system
per
unit length in the $x$ direction for various film thicknesses. A typical
behavior of 
$df_{el}^{(0)}/dh$ is shown in Fig.\ 1, where it is seen that
$f_{el}^{(0)}(h)$ indeed depends on the thickness $h$. Moreover, the
model predicts that
$d^{2}f_{el}^{(0)}/dh^{2}>0$, and therefore according to the inequality
(\ref
{crit}) and the discussion following it, there should be a linearly
stable wetting layer, whose thickness is finite and increases with
decreasing lattice mismatch.
\begin{figure}[h]
   \epsfxsize=75mm
   \centerline{\epsffile{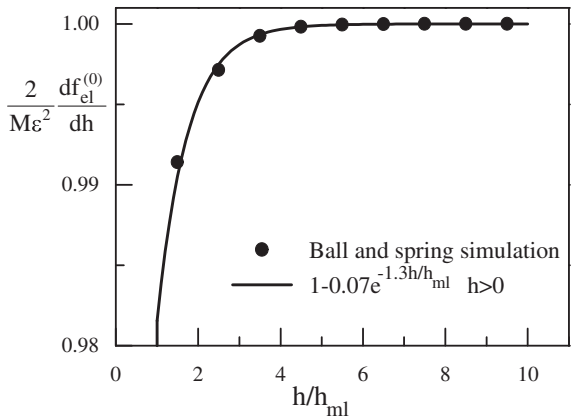}}
   \vspace{0.5cm}
\caption{Variation with film thickness of the elastic free energy of a
relaxed balls and springs system, $df_{el}^{(0)}/dh$, as a function of
film thickness $h$. The free energy is normalized to the 
infinite film linear elastic
energy density, $\frac{1}{2}M\varepsilon ^{2}$. $h_{ml}$ is the
thickness of one monolayer.}
\end{figure}
The dependence of $f_{el}^{(0)}$ on $h$ arises within our model from the
elastic relaxation at the surface and its coupling to the relaxation at
the interface between the substrate and the film. A similar effect
should occur in real systems due to surface reconstruction, for example.

Our calculations indicate that the general qualitative behavior of the
mismatch stress and the elastic energy are not sensitive to the lattice
structures and the values of the spring constants. While the detailed
behavior close to the substrate-film interface ($\lesssim 3$ monolayers)
varied
between different networks, it showed the same general behavior. In all
systems $d^{2}f_{el}^{(0)}/dh^{2}$ showed exponential decay with a decay
length of
about a monolayer away from the interface. For the calculations used
later
in this paper we used the function
$df_{el}^{(0)}/dh=M\varepsilon ^{2}\left[ 1-0.05\exp
(-h/h_{ml})\right]/2$
for $h>0$ and $df_{el}^{(0)}/dh=0$ for $h\leq 0$. $h_{ml}$ is the
thickness of one monolayer. The factor
$M\varepsilon^{2}/2$ on the r.h.s.\ ensures that the above expression
becomes exact
for a film of infinite thickness. 

Combining this behavior of $df_{el}^{(0)}/dh$ with the
inequality (\ref{crit}), we obtained an expression for the linear
stability
wetting layer thickness, $h_c$:
\begin{equation}
h_{c}/h_{ml}=
\max\left\{1,ln\left[\widetilde{\gamma}_0/(40M\varepsilon^2
h_{ml})\right]
\right\}~.
\label{hc}
\end{equation}
Thus, the wetting layer thickness increases with
decreasing
lattice mismatch, as observed in experiments. 

In previous works \cite{chiu,kukta,spencer} on the physics of the
wetting
layer it was assumed that the reference state energy variation is a
smooth
function of $h$, mainly in order to avoid non-analyticities at the
interface. In contrast, our reference state energy variation behaves as
a step function of the surface height with a small correction. We have
shown that the non-analytic behavior at the interface is realistic and
that the
smooth elastic energies in \cite{kukta,chiu,spencer} are unphysical.
Tersoff 
\cite{tersoff} in effect calculated $df_{el}^{(0)}/dh$ via an effective
atomic potential for Si/Ge under 4\% lattice mismatch. However, he did
not address the stability of the flat film to small perturbations.
Nevertheless, as the function he obtained had 
$d^{2}f_{el}^{(0)}/dh^{2}>0$, a positive wetting layer thickness can be
predicted from his results.

In order to model the early evolution of faceted islands, and to study
the effect of an anisotropic form of surface tension
on
the wetting layer, we used the
cusped form of surface tension given by Bonzel and Preuss \cite{bonzel},
which
shows
faceting in a free crystal: $\gamma (\theta )=\gamma _{0}\left[ 1+\beta
\left| \sin (\pi
\theta /(2\theta _{0}))\right| \right] $, where $\beta \approx 0.05$ and
$%
\theta _{0}$ is the angle of maximum $\gamma $. The value of $\gamma
_{0}$
was taken as 1 J/m$^{2}$ in the substrate and about 75\% 
of that in
the film (as is the case for Si/Ge). This ensures a
wetting
layer of at least one monolayer. We considered a crystal which facets at
0$%
^{\circ },\pm 45^{\circ }$ and $\pm 90^{\circ }$ with $\theta _{0}=\pi
/8$.
The cusp gives rise to $\widetilde{\gamma }=\infty $ and hence all
faceted
surfaces will have an infinite linearly stable wetting layer. However, a
slight miscut of the low-index surface leads to a rounding of the cusp,
which can be described by 
\begin{equation}
\gamma (\theta )=\gamma_0\left(1+\beta \sqrt{\sin^2 (\frac{\pi }{2\theta
_{0}}\theta
)+G^{-2}}\right)~,  \label{surf ten}
\end{equation}
where, for example, $G=500$ corresponds to a miscut angle, $\Delta
\theta\approx 0.1^{\circ}$. 
 
According to Eq.\ (\ref{hc}), anisotropic surface tension greatly
enlarges
the linearly stable wetting layer thickness. Does this conclusion
survive
beyond linear stability analysis? When a linearly stable flat film is
perturbed strongly so that the surface orientation in some regions is
far
from the $\theta =0$ direction, the local surface stiffness in these
regions
is much smaller than the $\theta =0$ stiffness. This tends to
destabilize the linearly stable film. Indeed, we carried out
Monte Carlo simulations that showed that films thinner than the linear
wetting
layer were unstable to random perturbations greater than a certain
critical
amplitude (see Fig.\ 2). The linear elastic energy
was calculated by the method used by Spencer
and Meiron \cite{meiron}. Hence films thinner than the linear wetting
layer
thickness are {\em metastable}. When large perturbations were applied,
faceted
islands developed in the film, which underwent Ostwald ripening at later
stages
of the evolution.

\begin{figure}[h]
   \epsfxsize=75mm
   \centerline{\epsffile{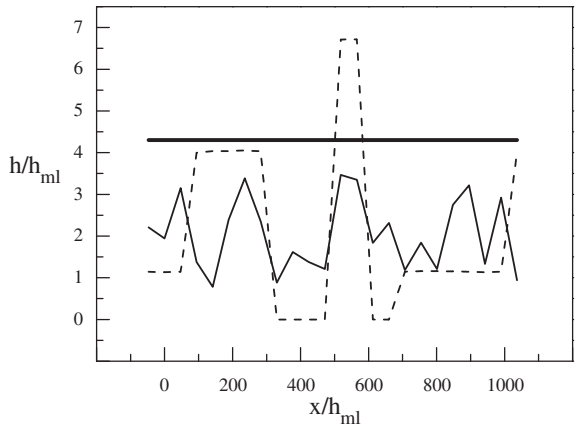}}
   \vspace{0.5cm}
\caption{Evolution of a randomly perturbed film, in which perturbations
were larger than the critical perturbation amplitude. Lattice mismatch
in this film is 4\%. The initial film surface is shown as a thin solid
line. The dashed line shows the film surface at a later time. The linear
wetting layer thickness is shown as a thick
solid line.}
\end{figure}

The critical perturbation amplitude was found to be proportional to
$\varepsilon ^{-2}$.
It
was largely independent of film thickness for $h<h_{c}$ as long as
perturbations did not penetrate the substrate, when it became much
larger.
It was also largely independent of cusp smoothness G, unlike the linear
wetting layer thickness which depended strongly on G. The size of the
critical perturbation amplitude in monolayers is plotted as a function
of
lattice mismatch in Fig. 3. The linear wetting layer thickness for
$G=500$, $M=1.5\times 10^{11} N/m^2$ and $h_{ml}=5{\mbox
\AA}$
is also shown for comparison. When the lattice mismatch is
small,
the critical perturbation amplitude is much larger than a monolayer.
Hence,
in practice,
flat films thinner than the linear critical thickness are stable at
small lattice mismatch. On the
other hand, for large mismatch a perturbation smaller than a monolayer
is
sufficient in order to destabilize the linearly stable wetting layer.
Therefore, in practice, the wetting layer will be a single monolayer at
large lattice mismatch. 
Our predictions cannot be compared with 
current experiments involving the wetting layer and lattice mismatch
variation \cite{floro}, since they were carried out with deposition
flux. We hope this work will
encourage such experiments to be performed, and we are currently adding
deposition to our model.

%
\begin{figure}[h]
   \epsfxsize=75mm
   \centerline{\epsffile{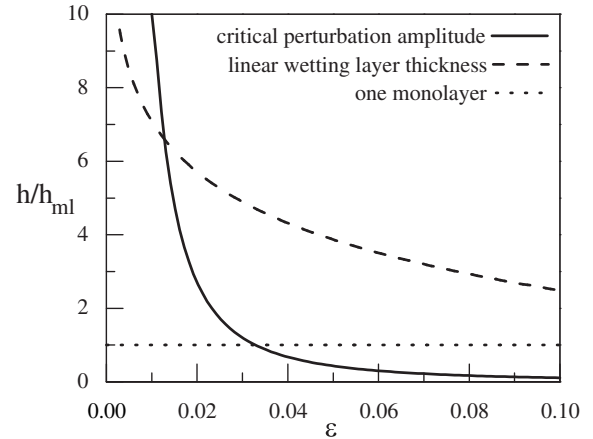}}
   \vspace{0.5cm}
\caption{Variation of critical perturbation amplitude with lattice
mismatch}
\end{figure}

This work was supported by the Israeli Science Foundation founded by
the Israeli Academy of Sciences and Humanities, and by the Israel
Ministry of Science.


\vspace*{-0.5cm}
\begin{thebibliography}{10}
\vspace*{-1.5cm}

\bibitem[*]{email1}
hrg1000@wicc.weizmann.ac.il.

\bibitem[**]{email2}
daniel.kandel@weizmann.ac.il, http://www.weizmann.ac.il/\~{ }fekandel.

\bibitem{mo}
Y.-W. Mo, D.E. Savage, B.S. Swartzentruber, M.G. Lagally, Phys. Rev.
Lett. {\bf
  65}, 1020 (1990); D.J. Eaglesham, M. Cerullo, Phys. Rev. Lett. {\bf
  64}, 1943
  (1990); J. Massies, N. Grandjean, Phys. Rev. Lett. {\bf 71}, 1411
  (1993).

\bibitem{floro}
J.A. Floro, E. Chason, R.D. Twesten, R.Q. Hwang, L.B. Freud, Phys. Rev.
Lett.
  {\bf 79}, 3946 (1997).

\bibitem{tersoff}
J. Tersoff, Phys. Rev. B. {\bf 43}, 9377 (1991).

\bibitem{spencer2}
B.J. Spencer, P.W. Voorhees and S.H. Davis, Phys. Rev. Lett. {\bf 67},
3696
  (1991).

\bibitem{chiu}
C.-H. Chiu, H. Gao, Mater. Res. Soc. Symp. Proc. {\bf 356}, 33 (1995).

\bibitem{kukta}
R.V. Kukta, L.B. Freund, J. Mech. Phys. Solids. {\bf 45}, 1835 (1997).

\bibitem{daruka}
I. Daruka and A.-L. Barab\'{a}si, Phys. Rev. Lett. {\bf 79}, 3708
(1997).

\bibitem{spencer}
B.J. Spencer, Phys. Rev. B, {\bf 59}, 2011 (1999).

\bibitem{timoshenko}
See, e.g., S. Timoshenko and J.N. Goodier, {\em Theory of Elasticity},
  McGraw-Hill (1951).

\bibitem{mullins57}
W.W. Mullins, J. Appl. Phys. {\bf 28}, 333 (1957); P. Nozi\`{e}res, J.
Physique
  {\bf 48}, 1605 (1987).

\bibitem{asaro}
R.J. Asaro, W.A. Tiller, Metall. Trans. {\bf 3}, 1789 (1972); D.J.
Srolovitz,
  Acta. Metall. {\bf 37}, 621 (1989).

\bibitem{herring}
C. Herring, Phys. Rev. {\bf 82}, 87 (1951).

\bibitem{bonzel}
H.P. Bonzel, E. Preuss, Surf. Sci. {\bf 336}, 209 (1995).

\bibitem{meiron}
B.J. Spencer, D.I. Meiron, Acta. Metall. Mater. {\bf 42}, 3629 (1994).

\end{thebibliography}

\end{document}